# LiDRoSIS: An Automated MATLAB–Python Platform for Image Processing and Quantitative Analysis of Lipid Droplets and ROS in Irradiated Cells

*Marco Ferreira* [1,2,3], *Ana Belchior* [2,4], *Teresa Pinheiro* [3,4], *Gil Alves* [2,5], *Maria Lopes* [2,6]

[1] Departamento de Física, Faculdade de Ciências, Universidade de Lisboa, Portugal
[2] Centro de Ciências e Tecnologias Nucleares, Instituto Superior Técnico, Portugal
[3] Institute of Bioengineering and Biosciences, Instituto Superior Técnico, Portugal
[4] Departamento de Engenharia e Ciências Nucleares, Instituto Superior Técnico, Universidade de Lisboa, Portugal
[5] Departamento de Física, NOVA School of Science and Technology, Universidade NOVA de Lisboa, Portugal
[6] Departamento de Física, Instituto Superior Técnico, Universidade de Lisboa, Portugal

**Abstract**
*LiDRoSIS is an automated MATLAB–Python software suite for the segmentation and quantification of lipid droplets (LDs) and reactive oxygen species (ROS) in fluorescence microscopy images of irradiated A549 and MCF7 cells exposed to gold-based nanoparticles. It combines classical image processing algorithms with statistical post-analysis through a companion Python tool, StatLysis™. The platform enables reproducible, high-throughput analysis of morphological and spectral parameters linked to nanoparticle-enhanced radiobiological responses. By bridging imaging and quantitative analytics, LiDRoSIS provides a robust framework for nanomedicine and radiation biology research.*

**Metadata**

| Nr | Code metadata description | *Metadata* |
|---|---|---|
| C1 | Current code version | *v1.0* |
| C2 | Permanent link to code/repository used for this code version | *https://github.com/marco18010104/LiDRoSIS* |
| C3 | Permanent link to reproducible capsule | *N/A- All scripts, example data, and usage instructions are provided in the public GitHub repository.* |
| C4 | Legal code license | *MIT License* |
| C5 | Code versioning system used | *git* |
| C6 | Software code languages, tools and services used | *MATLAB, Python* |
| C7 | Compilation requirements, operating environments and dependencies | *MATLAB Image Processing Toolbox; Python libraries (pandas, numpy, matplotlib, seaborn, scipy, statsmodels)* |
| C8 | If available, link to developer documentation/manual | *https://github.com/marco18010104/LiDRoSIS/blob/main/README.md* |
| C9 | Support email for questions | fc60327@alunos.fc.ul.pt |

# 1. Motivation and significance

*Microscopy-based quantification of subcellular features is central to understanding how ionizing radiation and nanoparticles (NPs) modulate cell physiology. However, segmentation of lipid droplets (LDs) and reactive oxygen species (ROS) in fluorescence microscopy (FM) images remains difficult due to illumination heterogeneity, morphological variability, and overlapping signals. Existing platforms such as ImageJ, CellProfiler, and 3D Slicer provide modular workflows but require manual intervention and lack integrated statistical post-processing.*

*__LiDRoSIS__ (Lipid Droplet and Reactive Oxygen Species Inference System) was developed to fill this gap. It is a standalone MATLAB application coupled to a Python analytics companion (StatLysis), designed to automate LD and ROS segmentation, quantification, and statistical evaluation in irradiated cell lines. The software provides reproducible, parameter-controlled processing and exports standardized outputs for cross-experiment comparisons.*

*The system addresses three key challenges in radiobiological imaging:*

1. ***Complex morphology:*** *radiation and nanoparticle exposure induce subtle morphological variations and diffuse fluorescence patterns in LD and ROS morphology that classical thresholding fails to capture.*
2. ***Reproducibility:*** *standardization of preprocessing and segmentation pipelines across large datasets.*
3. ***Scalability:*** *automation for batch processing of multiple fluorescence images.*

*LiDRoSIS implements intensity-based enhancement and morphological precision through a modular pipeline that integrates nuclei segmentation, blob detection (Difference of Gaussians and Steerable DoG), and unsupervised K-means clustering. It enables simultaneous analysis of red–green polarity shifts (via Nile Red staining) and ROS distribution (via DCFH-DA). The outputs are exported as Excel reports with three sheets - Global Metrics, Nucleus Metrics, and Object Metrics - which can be automatically aggregated and statistically analysed by StatLysis.*

*Beyond technical improvements, LiDRoSIS contributes to __scientific reproducibility__ in nanomedicine and radiobiology by providing an open, parameter-documented framework adaptable to different staining protocols and imaging conditions.*

*Comparable studies [Wen et al., 2023; Nurçin et al., 2022; Antunes et al., 2024] highlight the lack of specialized tools for irradiated cells with nanoparticles-LiDRoSIS explicitly targets this niche by combining classical methods and modern scripting flexibility without requiring deep learning expertise.*

*While open-source tools such as __CellProfiler__, __ImageJ/Fiji__, and __Ilastik__ provide general-purpose segmentation capabilities, they lack dedicated pipelines optimized for irradiated cellular models or fluorescence microscopy with overlapping vesicular structures. In radiobiological studies, the presence of nanoparticle-induced scattering, diffuse cytoplasmic signals, and overlapping emission spectra introduces substantial noise that conventional segmentation pipelines fail to handle robustly.*

*__LiDRoSIS__ addresses these gaps by combining adaptive morphological segmentation with reproducible post-analysis, bridging MATLAB's precision in image processing with Python's flexibility for statistical inference. The system was designed to ensure reproducibility and quantitative rigor in experiments involving __oxidative stress, radiation dose–response, and nanoparticle-enhanced radiosensitivity__.*

*In this context, LiDRoSIS contributes to the reproducible quantification of subcellular fluorescence dynamics - an increasingly critical component of biomedical imaging workflows, particularly in __radiation biology__, __toxicology__, and __nanomedicine__.*

## 2. Software description

*LiDRoSIS follows a hybrid design that merges classical image processing with statistical reproducibility, suitable for users without programming experience*

### 2.1. Software architecture:

LiDRoSIS comprises two main layers:

1. **Core segmentation (MATLAB):**
   - segmentNuclei: extracts nuclear masks using the DAPI (blue) channel with CLAHE enhancement and adaptive Otsu thresholding.
   - detectLDsRed, detectLDsGreen, and detectLDsColocalized: identify lipid droplets in red and green channels via DoG and SDOG filters, classify them by polarity, and compute colocalization indices.
   - detectDiffuseLDs: detects low-contrast lipid regions using K-means intensity clustering.
   - detectROS and detectDiffuseROS: identify punctate and diffuse oxidative structures from green-channel fluorescence.
   - All modules share nuclear masks to ensure biologically consistent per-cell assignment.
2. **Statistical analysis (Python):**
   - **StatLysis** imports the Excel outputs, aggregates metrics by experimental condition (cell line, nanoparticle type, radiation dose), and performs ANOVA, post-hoc tests, regression, and visualization (boxplots, violin plots, histograms).
   - It also generates model-fit summaries and plain-text conclusions for reproducible reporting.

The LiDRoSIS workflow follows a modular architecture that separates preprocessing, segmentation, feature extraction, and data export stages. Figure 1 summarizes the global pipeline.

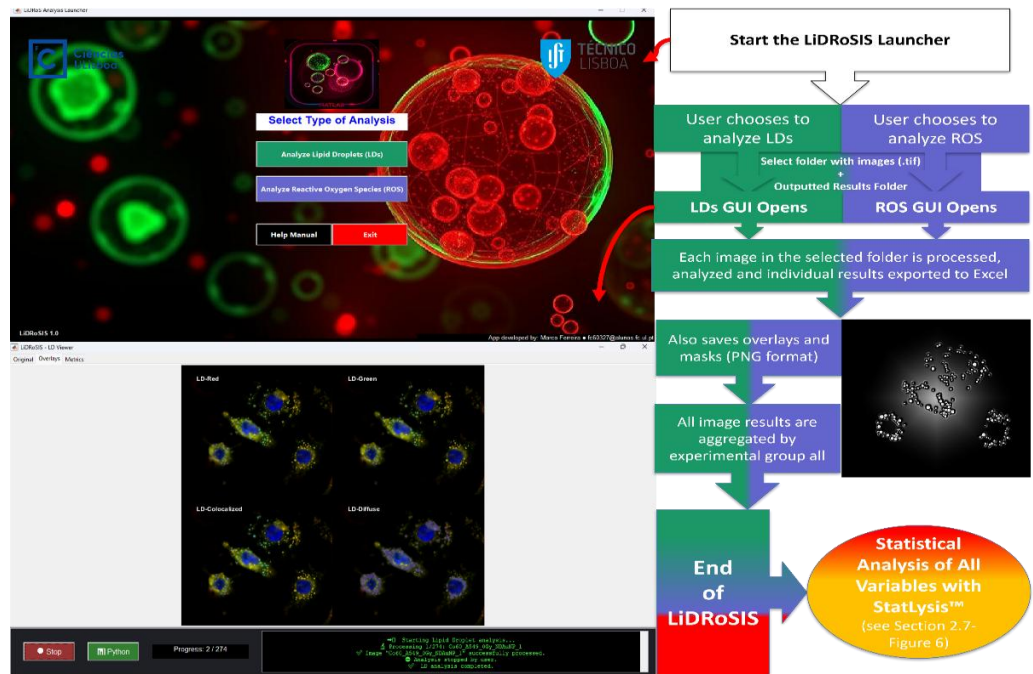

*Figure 1-Flowchart that provides a walkthrough of the program's functionality*

*The segmentation module employs adaptive intensity thresholding and morphological filtering tailored to fluorescence microscopy. For lipid droplets (LDs), the workflow includes Gaussian smoothing, Otsu-based thresholding, and circularity filtering to isolate near-spherical vesicles. For reactive oxygen species (ROS), segmentation relies on a combination of channel-specific intensity normalization and adaptive thresholding, allowing detection of both diffuse and punctate fluorescence patterns.*

*Following segmentation, each identified object is characterized by multiple metrics—including area, mean fluorescence intensity, circularity, eccentricity, and perimeter—which are stored per nucleus and globally.*

*The architecture supports interaction through GUI modules (guiLD.m, guiROS.m), enabling parameter tuning and batch processing. The main scripts (mainLD_gui.m, mainROS_gui.m) automatically export quantitative results to structured Excel sheets containing:*

- *GlobalMetrics – image-wide parameters (mean intensity, total droplet count, area),*
- *NucleusMetrics – data including normalized ROS intensity,*
- *ObjectMetrics – per-vesicle measurements.*

*A dedicated Python module, StatLysis.py, complements the MATLAB workflow by aggregating and statistically analysing the exported data. It supports ANOVA, linear regression, and boxplot visualization of ROS or LD metrics across experimental conditions. Integration between MATLAB and Python ensures full reproducibility from image input to statistical interpretation.*

## 2.2. Software functionalities:

*Major functionalities include:*

- ***Automated preprocessing:*** *color-channel separation, CLAHE contrast equalization, morphological filtering, and noise removal.*
- ***Blob-based segmentation:*** *multiscale DoG/SDOG filters to capture circular or elongated LDs and ROS foci.*
- ***Morphometric feature extraction:*** *area, eccentricity, solidity, circularity, perimeter, and equivalent diameter (using regionprops).*
- ***Biological validation:*** *nucleus-based assignment and exclusion of artefacts (e.g., edge nuclei, out-of-range intensity).*
- ***Batch analysis:*** *automatic processing of entire image folders.*
- ***Data export:*** *structured Excel reports with global, per-nucleus, and per-object statistics.*
- ***Statistical and graphical analysis:*** *automatic generation of violin plots, ANOVA tables, and dose-response regressions.*

*All parameters (thresholds, filters, size limits) are user-configurable through a GUI or a JSON configuration file, ensuring flexibility across different microscopy datasets.*

## 3. Illustrative examples

*To demonstrate LiDRoSIS performance, fluorescence microscopy images of **A549 (human lung carcinoma)** cells treated with gold-coated nanodiamonds (NDAuNPs) were analysed under three radiation doses (0 Gy, 2 Gy, and 10 Gy), with and without exposure to **gold nanoparticles (AuNPs)**. Figure 2 illustrates the segmentation of lipid droplets using the NileRed channel (red fluorescence). The algorithm effectively isolates vesicular structures while excluding diffuse cytoplasmic background. Quantitative metrics revealed a dose-dependent increase in droplet count and fluorescence intensity in irradiated and AuNP-treated cells.*

*Figure 3 presents an example of ROS segmentation results (green channel). The tool successfully distinguishes between localized vesicular ROS signals and broader cytoplasmic diffusion patterns; a limitation frequently encountered with conventional global thresholding.*

*Data exported from both modules were aggregated by dose and treatment condition using the Python companion script. Statistical boxplots and ANOVA tests confirmed significant fluorescence intensity variations ($p < 0.05$) between control and irradiated groups, validating the sensitivity of LiDRoSIS to radiation-induced oxidative stress.*

*LiDRoSIS successfully identified both red-shifted and colocalized LDs. The GlobalMeanRatio_Red_Green increased significantly with radiation dose (ANOVA, $p = 0.0011$), indicating LD polarity changes.*

*In MCF7 cells treated with AuNPs, LD abundance and red emission also increased with dose, while ROS intensity maps revealed enhanced oxidative stress. Outputs include binary masks, overlay visualizations, and quantitative summaries exported to Excel, as shown in Figure 2.*

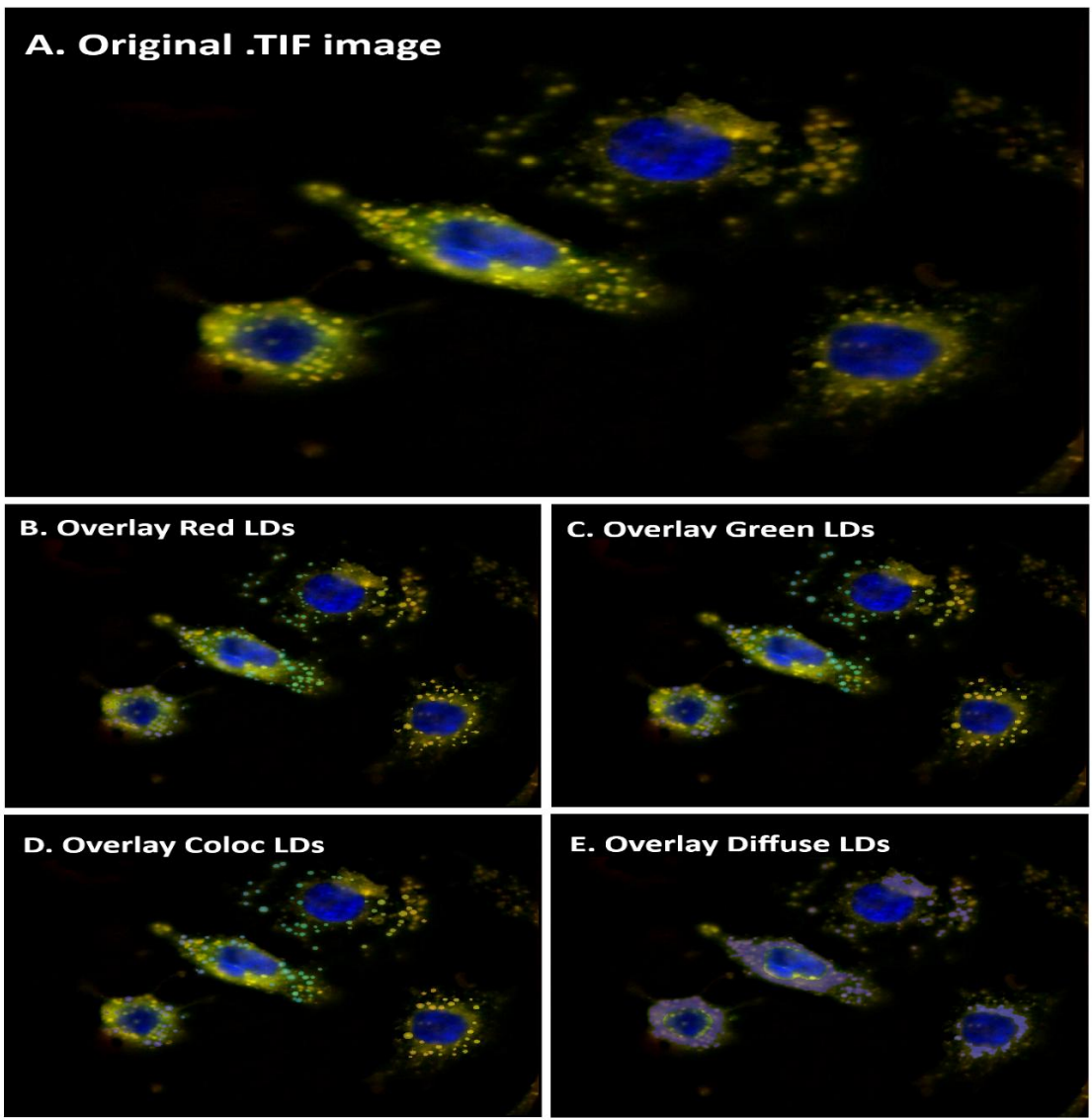

*Figure 2-Comparison between the (A) original .TIF image when inputted into the various functions that compose the LD algorithm. (B) Output image of detectLDsRed for (A); (C) Output image of detectLDsGreen for (A); (D) Output image of detectLDsColocalized for (A); (D) Output image of detectDiffuseLDs for (A). Example shown is of a FM image of A549 cells in the control group treated with NDAuNPs.*

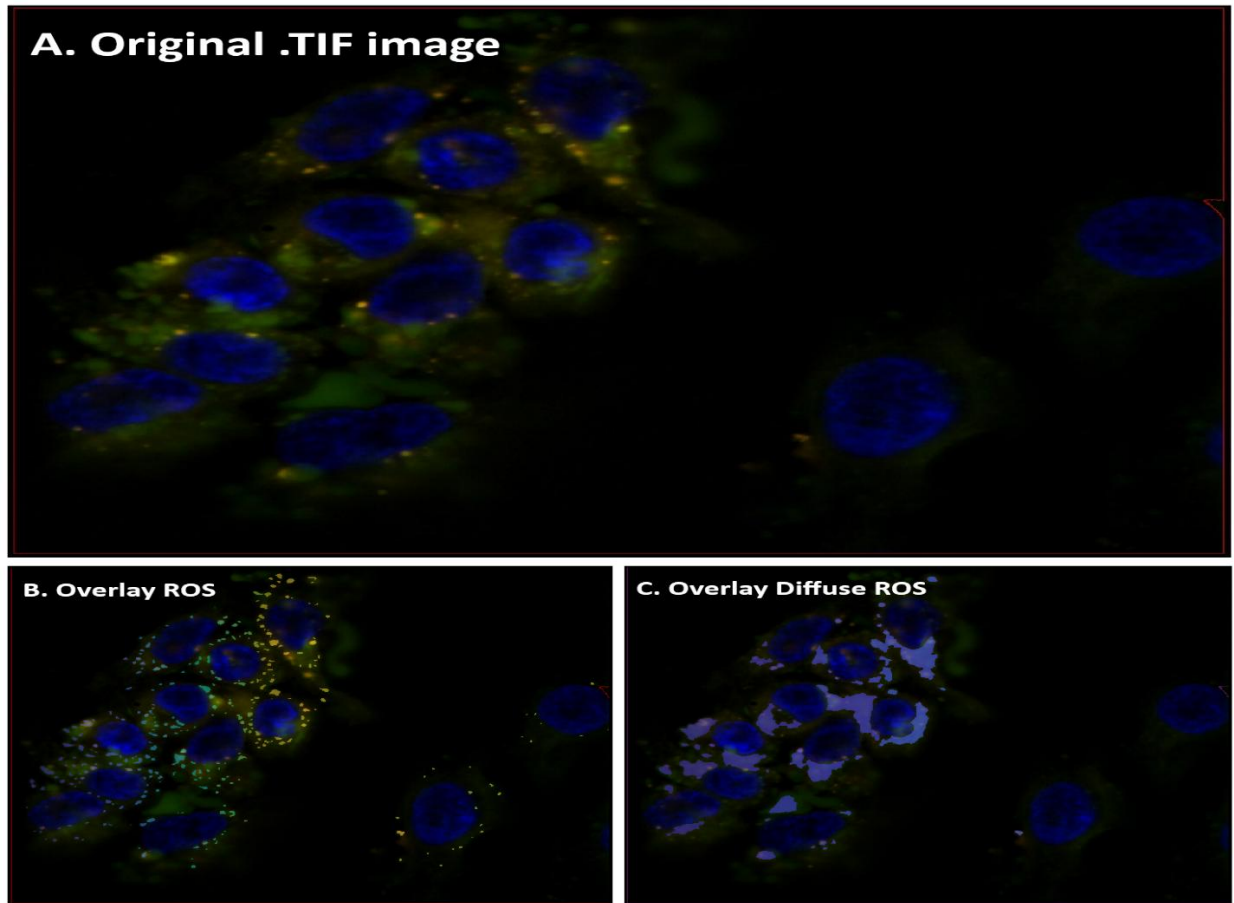

*Figure 3-Comparison between the (A) original .TIF image when inputted into the various functions that compose the ROS algorithm. (B) Output image of detectROS for (A); (C) Output image of detectDiffuseROS for (A). Example shown is of a FM image of A549 cells in the control group treated with NDAuNPs*

### 4. Impact

LiDRoSIS extends image analysis capacity in radiobiology by enabling:

- **Quantitative discovery:** new metrics linking lipid remodelling and oxidative stress to nanoparticle radiosensitization.
- **Improved reproducibility:** standardized image processing and metadata-preserving exports.
- **Accessibility:** runs on standard MATLAB/Python environments, usable without deep-learning frameworks.
- **Interdisciplinary use:** adaptable to any fluorescence microscopy data (cancer biology, toxicology, nanomaterials).
- **Research dissemination:** integration with Zenodo and GitHub ensures persistent accessibility and supports open-science and teaching applications.

Future research can build upon LiDRoSIS by integrating deep learning modules, 3D segmentation, or multimodal microscopy datasets.

Beyond its research utility, LiDRoSIS provides an educational framework for teaching image-based quantification and reproducible biomedical workflows. By bridging MATLAB and Python, the system illustrates how algorithmic transparency and open data practices can be combined in radiobiology research.

The open-source release under the MIT License enables straightforward extension to other imaging modalities, such as bright-field microscopy or hyperspectral fluorescence. Since its initial *deployment, LiDRoSIS has been downloaded and tested by researchers in cell biology and nanotoxicology laboratories, demonstrating its potential for community-driven adoption and expansion.*

**5. Conclusions**

*LiDRoSIS is an open, modular platform that automates fluorescence image segmentation of LDs and ROS in irradiated cells, combining classical, interpretable image processing with automated statistical analysis. Its reproducibility, transparency, and adaptability make it a valuable tool for radiobiology and nanomedicine research.*

**Acknowledgements**

*The author acknowledges Dr. Ana Lúcia Vital Belchior, Dr. Maria Teresa Pinheiro, and Prof. Nuno Matela for their supervision and support throughout the internship, as well as the Campus Tecnológico e Nuclear, Instituto Superior Técnico, for hosting the research internship.*